\documentclass[aps,prl,twocolumn,showpacs]{revtex4-1}
\usepackage{amsmath}
\usepackage{graphicx}
\usepackage{units}  
\usepackage{xspace}
\usepackage{subfigure}
\usepackage{hyperref}
\newcommand{\mb}{\mathbf}
\newcommand{\beq}{\begin{equation}}
\newcommand{\eeq}{\end{equation}}
\newcommand{\bea}{\begin{eqnarray}}
\newcommand{\eea}{\end{eqnarray}}

\newcommand{\moire}{moir\'{e}\xspace}

\usepackage[usenames]{color}

\begin{document}
\bibliographystyle{apsrev}
 
\title{Theory of Twisted Bilayer Graphene Near Commensuration}

\date{\today}
\author{Hridis K. Pal, Steven Carter, and M. Kindermann}
    \email{markus.kindermann@physics.gatech.edu}
\affiliation{School of Physics, Georgia Institute of Technology, Atlanta, GA 30332-0430, USA}

\begin{abstract}
Incommensurately twisted graphene bilayers are described by  long-wavelength theories, but to date such theories exist only at small angles of interlayer rotation. We construct a long wavelength theory without such a restriction, instead requiring nearness to  commensuration. The theory inherits its energy scale from the exactly commensurate bilayer that it is close to. It is a spatial interpolation between the low-energy theories of commensurate structures with the two possible sublattice exchange (SE) symmetries: SE even and SE odd. In addition to generalizing existing theories, our theory brings into experimental reach so far elusive commensuration physics in graphene such as band gaps and nontrivial band topology.
\end{abstract}

\maketitle 

In recent years there has been a surge of interest in   the properties of bilayer heterostructures with large moir\'{e} superlattices \cite{moire1,moire1a,moire1b,moire1c,moire1d,moire1e,moire1f,moire1g,moire1h,moire1i,moire2,moire3,moire4,moire5,moire6,moire7,moire8,moire9,moire10,moire11,moire12,moire13,moire14,moire15,moire16}. Such superlattices can occur in any bilayer system---twisted bilayer graphene, graphene on hexagonal boron nitride, bilayer dichalcogenides, etc.---as long as the two layers share the same crystal structure.
They originate from a mismatch between the two layers arising from either a relative twist, or a difference between the lattice constants, or a combination of both. Depending on the exact nature of the mismatch, the resulting structure could be commensurate or incommensurate. First-principles calculations of the electronic structure are feasible only in certain commensurate cases where the size of the supercell is not too large. Long-wavelength theories are, therefore, indispensable to the understanding of such structures. However, currently existing long-wavelength theories are restricted  to situations where the mismatch between the two layers is small. In incommensurately twisted bilayer graphene, for example, such theories exist only for small angles of rotation. 
 In this Letter, we construct a long wavelength theory without such a restriction, instead requiring nearness to  a commensurate structure. The theory facilitates the analysis of previously theoretically inaccessible types of  crystal bilayers which display  novel phenomena. 

As was shown by Lopes dos Santos, \emph{et al.} \cite{lopes:prl07},  at low energies the interlayer motion in incommensurate graphene bilayers with small angles of rotation is well approximated  by the lowest Fourier component of the interlayer coupling \cite{bistritzer:pnas11}. This leads to Dirac cones, as in single layer graphene, but with renormalized velocity. 
Subsequently, it was pointed out by Mele  \cite{mele:prb10} that  higher Fourier components of the interlayer coupling crucially enter the low-energy theory of commensurately rotated graphene bilayers.
They  induce a band curvature, and, in the case of even sublattice exchange (SE) symmetry, a gap in the spectrum. More recently it has been shown \cite{moire8}  that the bands in SE even  graphene bilayers are moreover topological: the material is a topological crystalline insulator. To date these intriguing aspects of commensuration physics have remained unobserved due to insufficient  control over the twist angle in experiments.

Here we  construct  a long-wavelength theory for twisted graphene bilayers at not necessarily small but  nearly commensurate angles. We find that the effects of the   higher Fourier components of the interlayer coupling previously studied   only at  commensurate angles  decide the physics  also at angles closeby. Besides extending the range of validity of long-wavelength theories for incommensurate bilayers, our theory thus greatly relaxes the experimental constraint on observing the above mentioned fascinating commensuration effects: They do not require exactly commensurate structures, but only rotation angles  within a  range of a commensurate one.
The width of this angular range depends on the theory at commensuration and it is largest near commensurations with  small  supercells. Our theory not only applies to incommensurate structures, but  also   to commensurate ones. In certain cases, the physics of a given commensurate structure is decided by   terms previously considered only at a nearby commensuration. Although we consider here explicitly the case of twisted bilayer graphene to construct our theory, the basic idea applies to any bilayer heterostructure with a moir\'{e} superlattice, irrespective of the cause of the moir\'{e} structure  or the underlying crystal symmetry.

\emph{Theory.}--- Consider a graphene bilayer with layers 1 and 2 rotated with respect to each other by an angle $\theta$. The Hamiltonian of the system  in the two-layer basis is
\beq
H=
\begin{pmatrix}
H_1&H_{\perp} \\
H_{\perp}^{\dagger}&H_2
\end{pmatrix},
\label{layerham}
\eeq
 where $H_{i}$ is the intralayer Hamiltonian of layer $i$  and $H_{\perp}$ couples the layers. 
In the continuum approximation one expands around the $K$ and $K'$ points. For ease of notation we give all expressions below only for K.
The individual layers are then described by Dirac Hamiltonians 
\beq
 H_{1\mb{k}}=v_F \boldsymbol{\sigma}_{ }\cdot\mathbf{k} ,\;\;\;H_{2\mb{k}}=v_F \boldsymbol{\sigma}_{ \theta}\cdot \mathbf{k} 
\label{dirac} 
\eeq
in momentum space, where $v_F$ is the Fermi velocity, $\boldsymbol{\sigma} = ( \sigma_x,\sigma_y)$ is a vector of Pauli matrices, and we set $\hbar=1$. Here and in what follows, a subscript $\theta$ on a vector denotes rotation by angle $\theta$.
The interlayer Hamiltonian $H_{\perp \mb{kk}'}$ in momentum space depends on the Fourier components $\tilde{t_{\perp}}(\mathbf{q})$ of the interlayer coupling $t_{\perp}(\mathbf{r})$.
It is convenient to make a gauge transformation of Eq.~(\ref{dirac}) that renders $H_1$ and $H_2$ identical at the expense of an extra $\theta$-dependence in $H_{\perp}$: $H_{\perp}\rightarrow H_{\perp}e^{-i \sigma_z\theta/2}$ with  \cite{lopes:prl07,mele:jphys12}
\begin{eqnarray}
H_{\perp \mb{kk}'}^{\alpha\beta}&=&\sum_{\mathbf{G},\mathbf{G}'_{\theta}}\tilde{t_{\perp}}(\mathbf{K}+\mathbf{k}+\mathbf{G}) e^{i(\mathbf{K}+\mathbf{G})\cdot\mathbf{\tau}^{\alpha}}e^{-i(\mathbf{K}_{\theta}+\mathbf{G'_{\theta}})\cdot\mathbf{\tau}_{\theta}^{\beta}}\nonumber\\
&&\delta(\mathbf{k}'-\mathbf{k}+\mathbf{K}_{\theta}-\mathbf{K}+\mathbf{G}'_{\theta}-\mathbf{G}). 
\label{hperp}
\end{eqnarray}
Here, $\alpha$ and $\beta$ are sublattice indices, $\mb{G}$ and $\mb{G}'$ are reciprocal lattice vectors, and $\mb{\tau}$ denotes the vector between the A and B atoms in a unit cell. 

For small angles of rotation, $|\theta|\ll 1$, $t_{\perp}(\mb{r})$ is a slowly varying function (on the lattice scale) and one can simplify Eq.~(\ref{hperp}) by considering only the  lowest Fourier components of $t_{\perp}(\mb{r})$. As first done by Lopes dos Santos \emph{et al.} \cite{lopes:prl07}, in this case one keeps  in Eq.\ (\ref{hperp}) the term with $\mb{G}=\mb{G}'_{\theta}=0$ and two more terms with reciprocal lattice vectors $\mb{G}'_{\theta}-\mb{G}$ which, when added to $\Delta\mathbf{K }=\mathbf{K}_{ \theta}-\mathbf{K} $, merely rotate the latter by  angle $2\pi n/3$ to $\Delta\mb{K}_{n}$, keeping $|\mathbf{K}_{ \theta}-\mathbf{K} +\mathbf{G}'_{\theta}-\mathbf{G}|=|\Delta\mb{K}_{n}|=|\Delta\mb{K} |$. All those three terms enter with a matrix element of magnitude $|\tilde{t}_{\perp}(\mb{K} )|\equiv\gamma/3$, where $\gamma$ is the nearest neighbor interlayer hopping in AA-stacked (or AB-stacked) bilayer graphene \cite{mele:prb10}. The momentum conservation condition encapsulated in the delta function in Eq.~(\ref{hperp}) reduces to $\delta({\mb{k}'-\mb{k} +\Delta\mb{K}_{  n}})$, so that each pair of states at wavevector $\mathbf{k}$ in one layer is coupled to three pairs of states at wavevectors $\mb{k} +\Delta\mb{K}_{n}$. This results in the preservation of the Dirac cones and the degeneracy at the Dirac point, only the velocity is renormalized. Mele developed the continuum theory further by including the effects of superlattice commensuration \cite{mele:prb10}.  The underlying idea is that if there is commensuration in real space, there is also commensuration in the reciprocal space. Therefore, in such commensurate cases there exist $\mb{G}$ and $\mb{G}'_{\theta}$ such that $\mb{K} +\mb{G}=\mb{K}_{ \theta}+\mb{G}'_{\theta}$, which, when inserted into Eq.~(\ref{hperp}), leads to the coupling of a state at wavevector $\mb{k} $ in one layer to a state at wavevector $\mb{k} $ in the other layer. The energy scale for this effect is governed by $\tilde{t}_{\perp}(\mb{K} +\mb{G})\equiv\mathcal{V}/3$, which decreases with $|\mb{K} +\mb{G}|$ and, therefore, with the size of the commensuration unit cell. The direct coupling between Dirac points induces deviations of the electronic spectrum from the massless Dirac form below the energy scale $\mathcal{V}$. 

\begin{figure}
\begin{center}
  \includegraphics[angle=0,width=0.6\columnwidth]{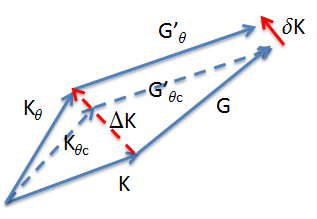}
  \caption{(Color online) Geometry of wacevectors illustrating the difference between $\Delta\mb{K} $ and $\delta\mb{K} $.}
\label{fig1}
\end{center}
\end{figure}

The theory by Lopes dos Santos \emph{et al.} is valid only in the small angle limit, when $\Delta K=2 K \mathrm{sin}\theta/2\ll K$. However,  by including higher Fourier components of the interlayer coupling, as done in Mele's theory at commensuration, one can construct a long-wavelength theory not just at small angles but at any  angle including $\theta\sim 1$, as long as it is near commensuration. To this end we consider a twisted graphene bilayer with an angle of rotation $\theta$, not necessarily small but only slightly away from a commensuration angle $\theta_c$, such that $|\delta\theta|=|\theta-\theta_c|\ll|\theta|$. 
 Since, by definition, $\theta_c$ leads to commensuration, we have $\mb{K} +\mb{G}=\mb{K}_{ \theta_c}+\mb{G}'_{\theta_c}$. Define $\mb{K}_{ \theta}+\mb{G}'_{\theta}-\mb{K} -\mb{G}=\delta\mb{K} $.
Clearly $\delta K\ll\Delta K$ since $|\delta\theta|\ll|\theta|$ and $\delta K\ll K$ since $|\delta\theta|\ll1$(cf. Fig.~\ref{fig1}). Our theory is built on the  observation that if, instead of expanding around the $ K$ point, we expand around $\mb{K} +\mb{G}$ in the extended Brillouin zone, $\delta\mb{K} $ appears naturally in the calculation instead of $\Delta\mb{K} $. Indeed, expanding around $\mb{K} +\mb{G}$ in one layer and $\mb{K}_{ \theta}+\mb{G}'_{\theta}$ in the other layer we find that the momentum conservation condition expressed by the delta function in Eq.~(\ref{hperp}) reduces to $\delta({\mb{k}'-\mb{k} +\delta\mb{K}_{  n}})$ as in the case of small angles, but with  $\Delta \mb{K}_{n}$ replaced by $\delta \mb{K}_{  n}$, the vector  $\delta \mb{K} $ rotated by $2\pi n/3$. Also, the intralayer Hamiltonians $H_1$ and $H_2$ remain unaltered as free Dirac Hamiltonians. Thus, it is possible to describe a system near commensuration, even for large rotation angles, by a theory similar to that of Ref.\ \cite{lopes:prl07},  but with reduced coupling   scale $\tilde{t}_{\perp}(\mb{K} +\mb{G})\equiv\mathcal{V}/3$ and reduced wavevector       $\delta\mb{K} $, instead of  $\tilde{t}_{\perp}(\mb{K} )\equiv\gamma/3$ and  $\Delta\mb{K} $, respectively. Expressing $\mb{G}$ and $\mb{G}'_{\theta_c}$ in terms of the reciprocal lattice vectors $\mathbf{b}_{1,2}=2\pi/3a_0(1,\pm\sqrt{3})$ of the graphene lattice ($a_0$ is the lattice constant) as $\mb{G}=l_1\mathbf{b}_1+l_2\mathbf{b}_2$ and $\mb{G}'_{\theta_c}=p_{1}\mathbf{b}_{1\theta_c}+p_{2}\mathbf{b}_{2\theta_c}$ such that $\mb{K} +\mb{G}=\mb{K}_{ \theta_c}+\mb{G}'_{\theta_c}$, we find the following expression for $H_\perp$:
\begin{equation}
H_{\perp}(\mb{r})=\frac{\mathcal{V}}{3}\sum_{n=0}^2e^{i \delta\mathbf{K}_n\cdot\mathbf{r}}
\begin{pmatrix}
1&e^{-i \frac{2\pi}{3}(n-p)}\\
e^{i \frac{2\pi}{3}(n-l)}&e^{-i \frac{2\pi}{3}(l-p)}
\end{pmatrix}
e^{-i \sigma_z\theta/2},
\label{hamreal}
\end{equation}
where $l=l_1+l_2$ and $p=p_1+p_1$. Note that as $\theta\rightarrow 0$, $e^{-i \sigma_z\theta/2}\approx 1$ and $\theta_c=0$ so that $\mathcal{V}=\gamma$, $l=p=0$, and $\delta\mb{K} =\Delta\mb{K} $. One thus recovers  \cite{kindermann:jphys12} the small angle theory of Ref.\   \cite{lopes:prl07} from Eq.~(\ref{hamreal}). Note that, although $H_{\perp}$ is written in terms of $\delta\theta$ (through $\delta\mb{K} $), information about the actual angle of rotation $\theta$ is still retained in two ways: implicitly through $\mathcal{V}$, $l$, and $p$, and explicitly through the term $\exp({-i \sigma_z\theta/2})$. Consequently, qualitatively new phenomena occur in incommensurate graphene bilayers at large angles, as discussed below. 

\emph{Discussion.}---The theory described above generalizes the existing long-wavelength theory of twisted graphene bilayers. The small angle theory of Ref.\  \cite{lopes:prl07}   is an important special case of Eq.\ (\ref{hamreal}), describing bilayers near the strongest commensuration at $\theta_c=0$. This implies that the previously discussed physics at small angles of rotation is not unique to small rotations but may also occur at large angles near commensuration, albeit at reduced energy scales.  For small angle rotations, $|\theta|\ll 1$, there are two regimes governed by $\gamma/v_F\Delta K$: a perturbative one, $\gamma/v_F\Delta K\ll 1$, where the Dirac spectrum is left intact as in single layer graphene with a renormalized velocity \cite{lopes:prl07}, and a nonperturbative one $\gamma/v_F\Delta K\gg 1$, where numerical calculations have pointed to the flattening of bands and localization \cite{laissardiere:nano10,bistritzer:pnas11}. Similar physics is expected for systems near  large angle commensurations with magnitude now governed by $\mathcal{V}/v_F\delta K$.  Note   that, although the coupling scale near large angle commensurations is typically substantially smaller than in the small angle case, the nonperturbative limit is always   reached at angles sufficiently close to commensuration, with sufficiently small $\delta K$.

The appearance of an increased length scale  $1/\delta\mb{K} $   in lieu of $1/\Delta\mb{K} $ in Eq.~(\ref{hamreal}) has an  intriguing consequence: it is not possible to uniquely determine the angle of interlayer rotation from the Moire pattern observed in scanning tunneling microscopy (STM) images based on the periodicity of the pattern alone, as is routinely done to date. To illustrate this we calculate the spatial dependence of the density of states (DOS) $\rho$ using the Hamiltonian  Eq.\ (\ref{hamreal}), perturbatively, to leading order in $\mathcal{V}/(v_F \delta K)$  \cite{kindermann:jphys12}:
\beq
\frac{\delta\rho(\mb{r})}{\rho_0}=f(\theta)\left(\frac{\mathcal{V}}{v_F\delta K}\right)^2\sum_{n\ne n'}\mathrm{cos}[(\delta\mathbf{K}_n-\delta\mathbf{K}_{n'})\cdot \mathbf{r})],
\label{DOS}
\eeq
where $f(\theta)=(2/9) \mathrm{cos}(\phi+2 \pi p/3) \mathrm{cos}(\phi+2 \pi l/3-\theta)$,
$\phi$ is the angle $\delta\mb{K} $ makes with $\mb{b}_1+\mb{b}_2$, and $n$, $n'$ run from 0 to 2. Being entirely determined by $\delta\mathbf{K}$ and not $\Delta\mathbf{K}$, the periodicity in Eq.~(\ref{DOS}) does not uniquely determine the absolute rotation angle $\theta$, but only its deviation from commensuration $\delta \theta$.  The actual angle of rotation  enters Eq.~(\ref{DOS}) only as a prefactor,  deciding the amplitude of the oscillations, but not their wavelength. 

Despite the similarities of our theory in Eq.~(\ref{hamreal}) with the small angle theory, the differences between the two extend  beyond a mere rescaling of length and energy scales, and have qualitative consequences. For example,
 in the nonperturbative limit ${\cal V}/v_F\delta K\gg 1$ the theory of Eq.\ (\ref{hamreal})   
 predicts local gaps due to the term $\exp({-i \sigma_z\theta/2})$. In fact,  our near commensurate theory is a spatial interpolation between regions where it locally takes the form of exactly commensurate structures:  SE even in regions that correspond to AA-stacking at small angles---a  topologically nontrivial gap arises in such regions---and SE odd in regions that are Bernal stacked in the small angle case---there are no gaps  in such regions.

Our theory thus greatly facilitates the experimental verification of the effects of commensuration as  predicted in Refs.\ \cite{mele:prb10, moire8}, hitherto unobserved experimentally due to insufficient experimental control over twist angles:
since the physics at commensuration also decides the physics near it, our theory opens a large window of angles at which experiments may be performed to verify commensuration effects.  That angular range is on the order of several degrees for simple manifestations of near commensuration physics, such as the long-wavelength density modulations of Eq.\ (\ref{DOS}). Observation of more intriguing effects such as gaps, or nontrivial band topology, pose more stringent conditions on the rotation angle: it then has to be within a range of  $\lesssim {\cal V}/v_FK$ radians. For instance, near the commensuration at $\theta_c=38.21^\circ$ \cite{mele:prb10}, this angular range is on the order of a tenth of a degree. It is expected to be increased by many-body effects   \cite{moire6}. 

A few words on the regime of applicability of our theory are in order.  By dint of the construction of the theory, in general,
the coupling term $\mathcal{V}$ is not the lowest Fourier component of $t_\perp$, unlike in the case of small angles. In the expansion of the interlayer coupling, there are other terms due to  lower (and thus larger) Fourier components. However, in  the perturbative regime, the effects are governed by the parameter  $\tilde{t}_{\perp}(\mb{K} +\mb{G})/|\mathbf{K}_{ \theta}-\mathbf{K}+\mathbf{G}'_{\theta}-\mathbf{G}|$ and sufficiently close to commensuration, $\mathcal{V}/v_F\delta K$ dominates over all other terms because of the smallness of $\delta K$. Moreover, at angles $\theta \simeq1$ all but possibly the term captured by Eq.\ (\ref{hamreal}) are indeed perturbative. Our theory, therefore, provides a good description of a system near commensuration as long as $\mathcal{V}/v_F\delta K$  dominates over terms due to other Fourier components. 
This answers any misgivings about the uniqueness of the theory for a given angle of rotation, since there can be, in principle, more than one angle of commensuration to which the system is close:
While the full theory is a sum over all commensurations,   the theory presented here captures the dominant contribution. Similarly, our theory  also applies to commensurate structures. For a commensurate angle  close to another one with a relatively small associated supercell,   our theory may dominate the physics at intermediate energies, and,  the term Eq.\   (\ref{hamreal}) needs to be included in the long-wavelength theory of the system.

We would like to point out that our theory can easily be  extended to other bilayer heterostructures with  moir\'{e} superlattices. The precise form of $H$ may change, but the essential idea that the system admits a long-wavelenth description near commensuration holds true irrespective of the cause of the mismatch between the two layers or even the underlying crystal structure. 

\begin{figure}
\begin{center}
  \includegraphics[angle=0,width=0.5\columnwidth]{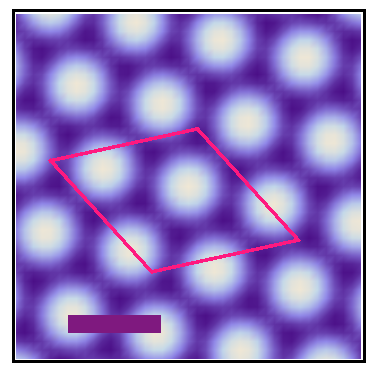}
  \caption{(Color online) Numerically calculated DOS at $\theta=35.57^\circ$. The interlayer coupling is   $V(\mathbf{r}_i,\mathbf{r}_j)=t_0 e^{-(|\mathbf{r}_i-\mathbf{r}_j|/l_0)^2}$ with $t_0=0.05t$ and $l_0=0.70a_0$, $t$ and $a_0$ being the in-plane hopping parameter and lattice constant, respectively. The rhombus denotes the supercell. The scale bar corresponds to $2 \pi/\sqrt{3}\delta K$.}
\label{fig2}
\end{center}
\end{figure}  

\emph{Tight-binding calculations.}---The veracity of our claims can be easily tested numerically. To this end we have performed    tight-binding calculations for a twisted graphene bilayer with interlayer rotation angle $\theta=35.57^\circ$. This angle is close to the commensurate rotation with   $\theta_c=38.21^\circ$, where the effects of commensuration are predicted to be the greatest  \cite{shallcross:prl08}. Such a choice yields $l=1$ and $p=-1$ in Eq.~(\ref{hamreal}).   We choose the interlayer hopping, for simplicity, to be a Gaussian: $V(\mathbf{r}_i,\mathbf{r}_j)=t_0 e^{-(|\mathbf{r}_i-\mathbf{r}_j|/l_0)^2}$. For the reference structure at $\theta_c$ we find for this interlayer coupling at $l_0=0.70 a_0$  a gap of $0.09t_0$ in the spectrum.  As expected, no discernible gap is found at the Dirac point for the twist angle $\theta$.

\begin{figure}
\begin{center}
  \includegraphics[angle=0,width=0.99\columnwidth]{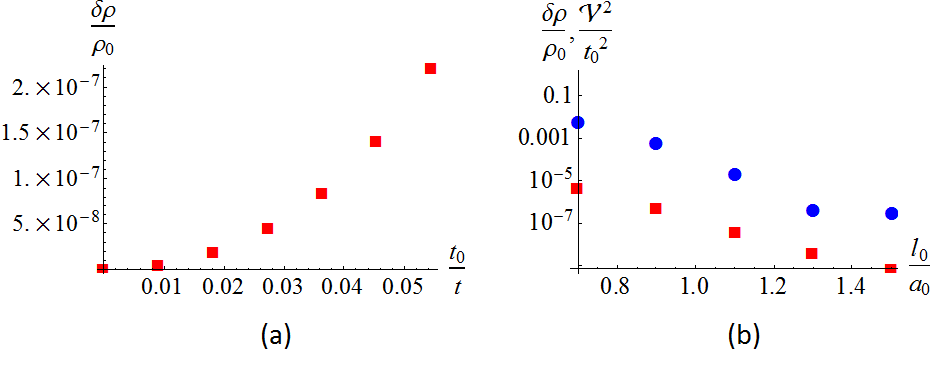}
  \caption{(Color online) (a) Dependence of $\delta\rho/\rho_0$ on $t_0$ (in units of in-plane hopping parameter $t$). (b) Dependence of $\delta\rho/\rho_0$ (square, red) and $\mathcal{V}^2$  (circle, blue) on $l_0$ (in units of lattice constant $a_0$). Note the logarithmic scale on the y-axis.}
\label{fig3}
\end{center}
\end{figure}

In Fig.~\ref{fig2} we plot the numerically calculated DOS at angle $\theta$. 
Density oscillations with wavevector $\delta\mb{K} $ and trigonal symmetry are clearly observed, as predicted by our long-wavelength theory, cf.\ Eq.~(\ref{DOS}).  
For further confirmation we study the dependence of the DOS on the parameters $t_0$ and $l_0$ in the interlayer coupling potential. Since $\mathcal{V}\propto t_0$, and from Eq.~(\ref{DOS}) $\delta\rho(\mathbf{r})/\rho_0\propto \mathcal{V}^2$, we expect the amplitude of the DOS oscillations  to vary quadratically with $t_0$, which is confirmed numerically in Fig.~\ref{fig3}(a). Similarly, increasing $l_0$ implies an   interlayer coupling with a smoother space-dependence. One expects the effects of commensuration governed by $\mathcal{V}$ to decrease quickly with increasing $l_0$,  since higher Fourier coefficients of the chosen interlayer coupling decay exponentially with $l_0$.  The corresponding decrease of the density oscillations $\delta\rho(\mathbf{r})/\rho_0\propto \mathcal{V}^2$ is clearly borne out   in  Fig.~\ref{fig3}(b).  A final check exploits the fact that the band splitting at the Dirac point in the commensurate case  is equal to $2\mathcal{V}$ \cite{mele:prb10}.  A tight-binding calculation of the spectrum at $\theta_c$ thus allows us to determine the value of $\mathcal{V}$. 
Since in our theory $\delta\rho(\mathbf{r})/\rho_0$  depends on $l_0$ only through $\mathcal{V}$, it is expected to depend on $l_0$ in the same way as $\mathcal{V}^2$ does. We confirm this in Fig.~\ref{fig3}(b): the $l_0$-dependence of $\mathcal{V}^2$, as found from the commensurate structure at  $\theta_c$,  indeed matches that of $\delta\rho(\mathbf{r})/\rho_0$ at angle $\theta$.  Moreover, using the value of  $\mathcal{V}$ found from the bilayer at $\theta_c$ in Eq.~(\ref{DOS})  one  predicts the amplitude of density oscillations without any  more free parameters.  We find that the numerically calculated DOS as shown in Fig.~\ref{fig2} indeed agrees with this prediction  to the precision of our theory, $\delta\theta/\theta_c$.

\emph{Conclusion.}---We have formulated a  long-wavelength theory for twisted graphene bilayers close to commensuration.  
The theory provides a unified long-wavelength description of such  bilayers in both small and large angle limits, thus generalizing previous theories \cite{lopes:prl07} valid only in the former limit. The theory has important ramifications for the extraction of the interlayer rotation angle from  \moire patterns seen in  STM.
It moreover predicts novel phenomena in graphene bilayers near large angle commensurations, such as local gaps.
 Intriguing effects had been predicted for some exactly commensurate graphene bilayers, such as  spectral gaps \cite{mele:prb10} and topological states  \cite{moire8} which, however, have remained unobserved  to date  due to the lack of experimental control over twist angles. Our theory greatly relaxes the experimental requirements,  
facilitating the observation of such effects not only at an exactly commensurate angle, but  in an entire angular range around it.   
Although we have formulated our theory for twisted bilayer graphene, the theory readily generalizes to other bilayer systems with moir\'e superlattices.

\begin{acknowledgements}
We thank J. Kunc for discussions and  acknowledge support by NSF under DMR-1055799.
\end{acknowledgements}


\begin{thebibliography}{99}


\bibitem{moire1} E. J. Mele, Phys. Rev. B \textbf{84}, 235439 (2011).

\bibitem{moire1a} C. Berger, Z. Song, X. Li, X Wu, N. Brown, C. Naud, D. Mayou, T. Li, J. Hass, A. N. Marchenkov, E. H. Conrad, P. N. First, W. A. de Heer, Science \textbf{312}, 1191 (2006).

\bibitem{moire1b} S. Latil, V. Meunier, and L. Henrard, Phys. Rev. B \textbf{76}, 201402(R) (2007).

\bibitem{moire1c} J. Hass, F. Varchon, J. E. Mill\'{a}n-Otoya, M Sprinkle, N. Sharma, W. A. de Heer, C. Berger, P. N. First, L. Magaud, and E. H. Conrad, Phys. Rev. Lett. \textbf{100}, 125504 (2008).

\bibitem{moire1d} M. Sprinkle, D. Siegel, Y. Hu, J. Hicks, A. Tejeda, A. Taleb-Ibrahimi, P. Le F\`{e}vre, F. Bertran, S. Vizzini, H. Enriquez, S. Chiang, P. Soukiassian, C. Berger, W. A. de Heer, A. Lanzara, and E. H. Conrad, Phys. Rev. Lett \textbf{103}, 226803 (2009).

\bibitem{moire1e} D. M. Miller, K. D. Kubista, G. M. Rutter, W. A. de Heer, P. N. First, and J. A. Stroscio, Science \textbf{324}, 9242 (2009).

\bibitem{moire1f} J. Hicks, M. Sprinkle, K. Shepperd, F. Wang, A. Tejeda, A. Taleb-Ibrahimi, F. Bertran, P. Le F\`{e}vre, W. A. de Heer,  C. Berger, and E. H. Conrad, Phys. Rev. B \textbf{83}, 205403 (2011).

\bibitem{moire1g} G. Li, A. Luican, and E. Y. Andrei, Phys. Rev Lett. \textbf{102}, 176804 (2009).

\bibitem{moire1h} G. Li, A. Luican, J. M. B. Lopes dos Santos, A. H. Castro Neto, A. Reina, J. Kong, and E. Y. Andrei, Nature Phys. \textbf{6}, 109 (2010).

\bibitem{moire1i} A. Luican, G. Li, A. Reina, J. Kong, R. R. Nair, K. S. Novoselov, A. K. Geim, and E. Y. Andrei, Phys. Rev. Lett \textbf{106}, 126802 (2011).

\bibitem{moire2} A. Pal and E. J. Mele, Phys. Rev. B \textbf{87}, 205444 (2013).

\bibitem{moire3} M. Diez, J. P. Dahlhaus, M. Wimmer, and C. W. J. Beenakker, Phys. Rev. Lett. \textbf{112}, 196602 (2014).

\bibitem{moire4} X. Chen, J. R. Wallbank, A. A. Patel, M. Mucha-Kruczy\'{n}ski, E. McCann, and V. I. Fal'ko, Phys. Rev. B \textbf{89}, 075401 (2014).

\bibitem{moire5} P. San-Jose, \'{A}. Guti\'{e}rrez, M. Sturla, and F. Guinea, arXiv:1404.7777.

\bibitem{moire6} J. C. W. Song, A. V. Shytov, and L. S. Levitov, Phys. Rev. Lett. \textbf{111}, 266801 (2013).

\bibitem{moire7} L. A. Ponomarenko,	R. V. Gorbachev,  G. L. Yu,	D. C. Elias,	R. Jalil,	A. A. Patel,	A. Mishchenko,	A. S. Mayorov,	C. R. Woods,	J. R. Wallbank,	M. Mucha-Kruczynski,	B. A. Piot, M. Potemski,	I. V. Grigorieva,	K. S. Novoselov, 	F. Guinea,	V. I. Fal’ko, and A. K. Geim, Nature \textbf{497}, 594 (2013).

\bibitem{moire8} M. Kindermann, arXiv:1309.1667.

\bibitem{moire9} M. Yankowitz,	J. Xue,	D. Cormode,	J. D. Sanchez-Yamagishi,	K. Watanabe,	T. Taniguchi,	P. Jarillo-Herrero,	P. Jacquod, and B. J. LeRoy, Nature Phys. \textbf{8}, 382 (2012).

\bibitem{moire10} G. L. Yu, R. V. Gorbachev, J. S. Tu, A. V. Kretinin, Y. Cao, R. Jalil, F. Withers, L. A. Ponomarenko, B. A. Piot, M. Potemski, D. C. Elias, X. Chen, K. Watanabe, T. Taniguchi, I. V. Grigorieva, K. S. Novoselov, V. I. Fal'ko, A. K. Geim, and A. Mishchenko, Nature Phys. \textbf{10}, 525 (2014).

\bibitem{moire11} C. R. Woods,	 L. Britnell,	A. Eckmann,	R. S. Ma,	J. C. Lu,	H. M. Guo,	X. Lin,	G. L. Yu,	Y. Cao, R. V. Gorbachev,	A. V. Kretinin,	J. Park,	L. A. Ponomarenko,	M. I. Katsnelson,	Yu. N. Gornostyrev,	K. Watanabe,	T. Taniguchi,	C. Casiraghi,	H-J. Gao,	A. K. Geim, and K. S. Novoselov, Nature Phys. \textbf{10}, 451 (2014).

\bibitem{moire12} M. Kindermann, B. Uchoa, and D. L. Miller, Phys. Rev. B \textbf{86}, 115415 (2012).

\bibitem{moire13} P. San-Jose, J. Gonzalez, and F. Guinea, Phys. Rev. Lett. \textbf{108}, 216802 (2012).

\bibitem{moire14} J. Kang,  J. Li, S. Li, J. Xia, and L. Wang, Nano Lett. \textbf{13}, 5485 (2013).

\bibitem{moire15} A. K. Geim and I. V. Grigorieva, Nature \textbf{499}, 419 (2013). 

\bibitem{moire16} K. Liu, L. Zhang, T. Cao, C. Jin, D. Qiu, Q. Zhou, A. Zettl, P. Yang, S. G. Louie, and F. Wang, arXiv:1406.6487.

\bibitem{lopes:prl07} J. M. B. Lopes dos Santos, N. M. R. Peres, and A. H. Castro Neto, Phys. Rev. Lett. \textbf{99}, 256802 (2007).

\bibitem{bistritzer:pnas11} R. Bistritzer and A. H. MacDonald, PNAS \textbf{108}, 12233 (2011).

\bibitem{mele:prb10} E. J. Mele, Phys. Rev. B \textbf{81}, 161405(R) (2010).

\bibitem{mele:jphys12} E. J. Mele, J. Phys. D: Appl. Phys. \textbf{45}, 154004 (2012).

\bibitem{kindermann:jphys12} M. Kindermann and P. N. First, J. Phys. D: Appl. Phys. \textbf{45}, 154005 (2012).

\bibitem{laissardiere:nano10} G. Trambly de Laissardie\`{r}e, D. Mayou, and L. Magaud, Nano Lett. \textbf{10}, 804 (2010).

\bibitem{shallcross:prl08} S. Shallcross, S. Sharma, and O. A. Pankratov, Phys. Rev. Lett. \textbf{101}, 056803 (2008).


\end{thebibliography}
\end{document}